\documentclass[aps,prl,twocolumn,showpacs]{revtex4-1}
\usepackage{graphicx}
\usepackage{subfigure}
\usepackage{amsmath}
\usepackage{epsfig}
\usepackage{epstopdf}
\usepackage{color}
\usepackage{hyperref}
\usepackage{tikz}

\usepackage[T1]{fontenc}

\begin{document} 
	\title{Random walks with asymmetric time delays} 
	\author{Kamil \L{}opusza\'{n}ski}
	\affiliation{Faculty of Mathematics, Informatics and Mechanics \\ University of Warsaw \\ ul. Banacha 2 \\
		02-097 Warsaw, Poland \\ email: kamil.lopuszanski@gmail.com}
	\author{Jacek Mi\k{e}kisz}
	\affiliation{Institute of Applied Mathematics and Mechanics \\ University of Warsaw  \\ ul. Banacha 2  \\ 02-097 Warsaw, Poland \\ email: miekisz@mimuw.edu.pl}
	
	\begin{abstract}
		
We studied simple random-walk models with asymmetric time delays. Stochastic simulations were performed for hyperbolic-tangent fitness functions and to obtain analytical results we approximated them by step functions. A novel behavior has been observed. Namely, the mean position of a walker depends on time delays. This is a joint effect of both stochasticity and time delays present in the system. We also observed that by shifting appropriately fitness functions we may reverse the effect of time delays - the mean position of the walker changes the sign.

	\end{abstract}
	
	\maketitle
	
	\vspace{3mm}
	
	{\bf Introduction.} Many social and biological processes can be described by deterministic population dynamics  \cite{hofbook,nowakbook}.
	It is usually assumed that interactions between individuals take place instantaneously and their effects are immediate. 
	In reality, all processes take a certain amount of time. Results of biological interactions between individuals 
	may appear in the future, and in social models, individuals or players may act, that is choose appropriate strategies, 
	on the basis of the information concerning events in the past. It is natural therefore to introduce time delays to describe such processes. 
	It is well known that time delays may cause oscillations in dynamical systems \cite{ladas,gopalsamy,kuang,erneux}.
	One usually expects that interior equilibria of evolving populations - describing a coexistence of strategies or behaviors - are asymptotically stable for small time delays and above a critical time delay, where the Hopf bifurcation appears, they become unstable.
	
	Effects of time delays in replicator dynamics describing evolution of populations of individuals interacting through playing games \cite{taylor} were discussed in 
	\cite{tao,alboszta,aoku,ijima1,ijima2,moreira,wesol,matusz,wessonrand1,wessonrand2,nesrine1,nesrine2,jmmb1,jmmb2} 
	for games with an interior stable equilibrium (an evolutionarily stable strategy \cite{maynard2}). 
	There were studied models with strategy-dependent time delays recently. In particular, Moreira et al. \cite{moreira} discussed 
	multi-player Stag Hunt game with time delays, Ben Khalifa et al. \cite{nesrine1} investigated asymmetric games in interacting communities, Wesson and Rand \cite{wessonrand1} studied Hopf bifurcations in two-strategy delayed replicator dynamics. Systematic analysis of two player games with two strategies and strategy-dependent time delays is presented in \cite{jmmb2}. A novel behavior --- the continuous dependence of equilibria on time delays --- was observed.
	
	In finite populations, one has to take into account stochastic fluctuations \cite{nowakbook}. One of the simplest models involving both stochasticity and time delays is a delayed random walk \cite{ohira1,ohira2}. In such a walk, transition probabilities at any given time depend on the position of the walker at some earlier time. In \cite{ohira1}, a delayed random walk was considered, where in the absence of delays, 
	the transition toward the origin (a stable state) is more probable than the outward transition. 
	The authors show that the mean square displacement of the walker, that is the variance, approaches a stationary value 
	in an oscillatory manner for large time delays and in a monotonic way for small ones. Moreover, the stationary value of the variance is a linear function of the delay and the coefficient of the proportionality is a linear function of the transition probability. 
	
	Here we consider a modification of the above model. Namely, transition probabilities depend 
	on the difference between two functions (which can be interpreted as fitness functions in discrete replicator dynamics) 
	evaluated at different times in the past. To simplify our model, we set one of the time delays to zero. As fitness functions we use hyperbolic-tangent and step functions. In the absence of time delays and stochasticity, our models are discrete (in time and space) replicator-type dynamics of evolutionary games with an interior asymptotically stable stationary state.
	
	It should be noted here that such dynamics are not Markovian. However, we may restore Markovianity, if we consider transition probabilities not between states, but between histories of states. Such Markov chains are called higher-order Markov Chains \cite{higher}. 
	Although a stationary probability distribution is defined on histories, we will still be interested in the stationary probability of visiting particular states of the physical space. Our main object of study is the expected value of such a stationary probability distribution, its dependence on time delays and other parameters of our models. We have performed stochastic simulations for hyperbolic-tangent and step-function models and we have derived analytical formulas in the step-function case for some small time delays.
	
	We report a novel behavior - dependence of stationary probability distribution on time delays. This is a joint effect of stochasticity and time delays present in the system - in the deterministic version of our model, there appear symmetric cycles around the stationary point so the mean position of the walker stays the same, in the stochastic version without time delays, the expected value of the position of the walker is given by the stationary point. 
	
	{\bf Deterministic dynamics with time delays.} The state space $S$ of the walker is the set of all integers.
	We introduce two fitness functions on $S$, $f_{A}$ and $f_{B}$, such that $f_{A}$ is a decreasing function and $f_{B}$ is an increasing one,
	$f_{A}(0)=f_{B}(0)$, $f_{A} > f_{B}$ on $(-\infty,0)$ and $f_{A} < f_{B}$ on $(0,\infty)$. 
	Deterministic dynamics is given by the following rules: if $f_{A}(x(t-\tau)) - f_{B}(x(t)) > 0$, then the walker moves to the right at $t+1$, in the case of the reverse inequality, it moves to the left, in the case of the equality, it stays at a current position. We have to specify initial conditions. For systems with time delays it is a history \{$x(-\tau), x(-\tau +1), \ldots, x(-1), x(0)\}$.
	
	Obviously, the origin is a stationary state of such dynamics. However, it is unstable. 
	One can see that if the walker moves to the right or to the left of the origin, then there appears immediately a cycle around the origin, with the period $2\tau + 2$ and the amplitude $1$. It is also easy to see that if we start with initial conditions entirely on the right or on the left of the origin, then in a finite number of steps the system develops a cycle with the period $2\tau +2$ and the amplitude equal to the smallest integer bigger or equal to $\tau/2$. Therefore, after a finite number of steps, the system moves along a cycle, there are no other stationary states besides the origin.
	
	Now we will show that any such cycle has the period $2\tau + 2$, it is symmetric around the origin and therefore the average walker position along its trajectory is equal to $0$.
	
\begin{figure}
		\centering
		\includegraphics[width=\linewidth]{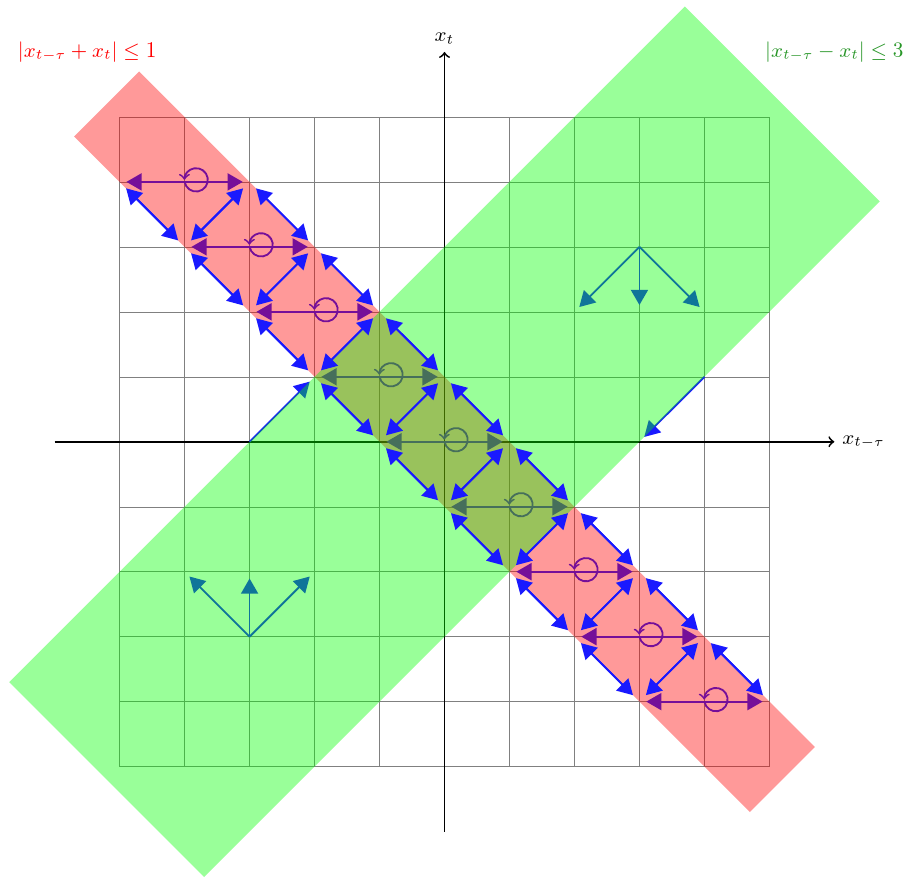}	
		
		\caption{Phase portrait in the phase space $(x_{t-\tau}, x_{t})$ for the deterministic dynamics. Blue arrows show directions in which the system can move. The state space $\Omega$ for $ \tau=3 $ is shown in green, the invariant set $T$ is shown in red.}
		\label{fig:deterministic-tangents-phase-portrait}
	\end{figure}

	We introduce a phase space of our system: $\Omega = \{(x(t-\tau), x(t)\}  = \{(m, n) \in Z^{2}, |m-n| \leq \tau\}$. Dynamics can be represented by walks over edges or diagonals of elementary squares of $\Omega$, see Fig. \ref{fig:deterministic-tangents-phase-portrait}. 
	Let $T=\{(m, n), |m+n| \leq 1\}$. It follows from the phase portrait in Fig. \ref{fig:deterministic-tangents-phase-portrait} that
	$T \cap \Omega$ is the attractor and the invariant set of the dynamics. Moreover, if the system moves on $T$ from $(i,j)$ to $(k,l)$ 
	in one step, then $l=-i$. Now we have the following path,
	\begin{equation}
		(i, j) \xrightarrow{1} (m, -i) \xrightarrow{\tau-1} (j, n) \xrightarrow{1} (-i, -j).
	\end{equation}
	
	Hence, every cycle is symmetric around the origin and the average position of the walker is $0$. 
	After $2\cdot(\tau+1) = 2\tau + 2$ steps, the system returns to the initial state. Hence any cycle must have a period which is a divisor of $2\tau + 2$.
	
	{\bf Stochastic dynamics with time delays.} Now we assume that a probability of the walker to move to the center, being at a position $x$, is proportional to the absolute value of the difference $f_{A}(x) - f_{B}(x)$ and such that a probability of moving to the center is higher than the one of moving outward. In this way we constructed an ergodic Markov chain with a unique stationary probability distribution symmetric 
	around the origin and with the expected value of the walker position equal to zero. It was shown in \cite{wesol}, 
	that when one introduces a discrete time delay $\tau$, that is transition probabilities depend on the difference  $f_{A}(x(t-\tau)) - f_{B}(x(t-\tau))$, then there appears a cycle around the origin with the amplitude $\tau$ and the time period $4\tau + 2$ which is stochastically stable \cite{freidlin,young,cime}. It means that the stationary probability is concentrated on the cycle with the probability converging to one in the limit of zero stochasticity. 
	
	Here we will consider asymmetric time delays --- we assume that transition probabilities depend on the difference $f_{A}(x(t-\tau)) - f_{B}(x(t))$. Formally our Markov chain is described by transition probabilities on the set of consistent histories 
	$\{x(t-\tau), x(t-\tau +1), \ldots , x(t-1), x(t)\}$.
	We define transition probabilities in the following way. When the walker is at state $x$, then the probability of moving to the right is given by
	\begin{equation}\label{eq:transition-prob}
		p_{+}(x(t)) = \frac{1}{2} + \omega \frac{f_{A}(x(t-\tau)) - f_{B}(x(t))}{2}, 
	\end{equation}
	where $\omega \leq 1$ is a noise parameter. Probability of moving to the left, $p_{-}(x(t))$, is given by $1- p_{+}(x(t))$. 
	
\begin{figure}
		\centering
		\includegraphics[width=\linewidth]{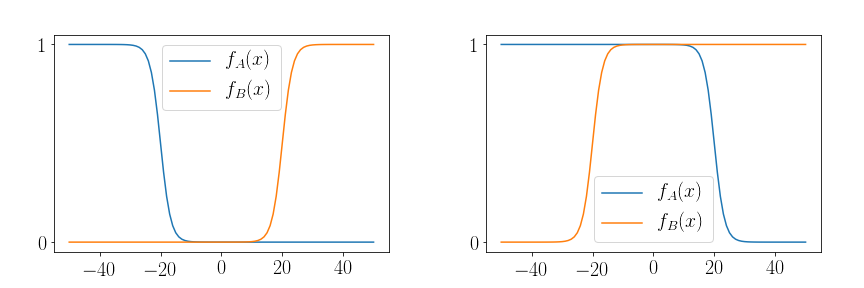}
		
		\caption{Hyperbolic tangent fitness functions; $d=-20$ on the left and $d=20$ on the right, $\rho=0.3.$}
		\label{fig:smooth}
	\end{figure}

	For fitness functions we use hyperbolic tangents: 
	\begin{subequations}
		\begin{align}
			\label{smooth_fa}
			f_A(x) &= \frac{1}{2}[\tanh(-\rho(x-d)) + 1], \\ 
			\label{smooth_fb}
			f_B(x) &= \frac{1}{2}[\tanh(\rho(x+d)) + 1],
		\end{align}
	\end{subequations}
	where \(|d|\) is the length of the half of the interval of an almost non-biased random walk and \(\rho\) is a
	coefficient. Fig. \ref{fig:smooth} shows plots of the fitness functions for
	\(d=-20, 20, \ \rho=0.3\).
	
	In this way, we defined a Markov chain on histories, a so-called higher-order Markov chain \cite{higher}, with a unique stationary probability distribution which we denote by $\pi^{h}$. Although states of our Markov chain are time histories, we are interested in frequencies in which particular physical states in $S$ are visited in the long-run --- in the stationary probability distribution. We denote by $\pi$ a stationary probability of visiting physical states in $S$. Both types of probabilities are related by simple total probability formulas as is shown later.
	
	\begin{figure}
		\centering
		\includegraphics[width=\linewidth]{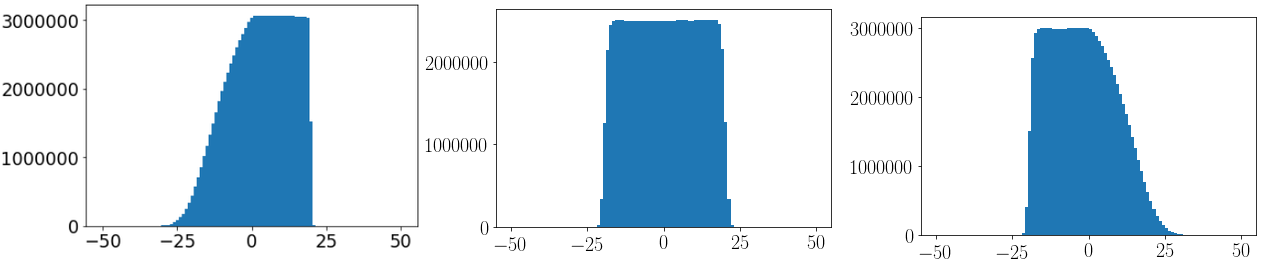}
		
		\caption{Histograms for the hyperbolic-tangent models with parameters $ \tau =20$ on the left and right, $ \tau_A=0$ in the middle. 
$ d=-20$ on the left and $ d=20$ on the right. $\rho=0.99 $ for all simulations. Simulation were run for $10^8$ steps.}
		\label{fig:histograms}
	\end{figure}
	
\begin{figure}
		\centering
		\includegraphics[width=\linewidth]{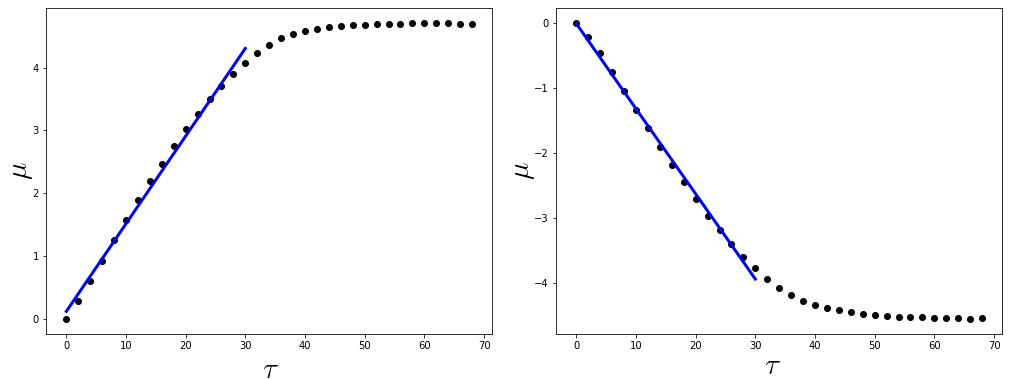}
		
		\caption{The dependence of the expected value of the walker's position $\mu$ on the time delay $\tau$ with hyperbolic-tangent fitness functions with $\rho=0.99$, $\tau=0, 2, 4, \ldots 68$. $d=-20$ on the left and $d=20$ on the right. The equation of the blue line (linear regression) is $0.14\tau + 0.12$ on the left and $ -0.13\tau - 0.01$ on the right.}
		\label{fig:tau-d20}
	\end{figure}

	We performed stochastic simulations of our dynamics, histograms of physical-state frequencies are presented in Fig. \ref{fig:histograms}. 
	One can see that histograms are skewed. 
	In Fig. \ref{fig:tau-d20}, we show the dependence of the expected value of the walker's position on the time delay.
	For a certain region of relatively small time delays, this dependence is given by a decreasing linear function for $d>0$ and an increasing one for $d<0$.
	By the use of a linear regression we approximated the slope of these functions to be $-0.13$ and $0.14$ respectively.
	
\begin{figure}
		\centering
		\includegraphics[width=\linewidth]{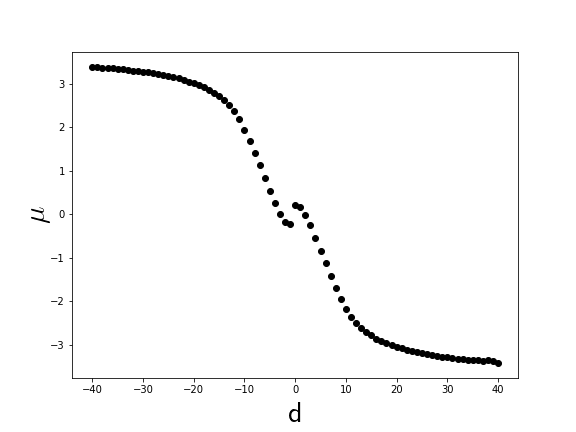}
		
		\caption{The dependence of the expected value of the walker's position $\mu$ on $d$ with hyperbolic-tangent fitness functions with  $\rho=0.99$, $\tau=20$, $d=-40, \ldots, -2, 0, 2, \ldots, 40$.}
		\label{fig:d-tau20}
	\end{figure}

	In Fig. \ref{fig:d-tau20}, we show the dependence of the expected value of the walker's position on the size of the space region with an almost unbiased random walk. We see that the expected value is a decreasing function of the size.
	
	Our results show a novel behavior not observed in any previous models. The shift of the expected value of the stationary probability distribution is a joint effect of both time delays and stochasticity. Moreover, when we shift the fitness functions, that is we change the sign of $d$, the effect of times delays is reversed.
	
	{\bf Analytical results.} Here we present some analytical results for fitness step functions which approximate hyperbolic tangents. 
	We set $\omega=1$, so that the only transition probabilities are $0,\frac{1}{2}$, and $1$. 
	Namely, let
	
	\begin{subequations}
		\begin{align}
			\label{eq:step-functions}
			f_A(x) &= \begin{cases}
				1 &\mbox{for } x \le d \\
				0 & \mbox{for } x > d
			\end{cases} \\
			f_B(x) &= \begin{cases}
				0 & \mbox{for } x < -d \\
				1 & \mbox{for } x \ge -d
			\end{cases}
		\end{align}
	\end{subequations}
	
\begin{figure}
		\centering
		\includegraphics[width=\linewidth]{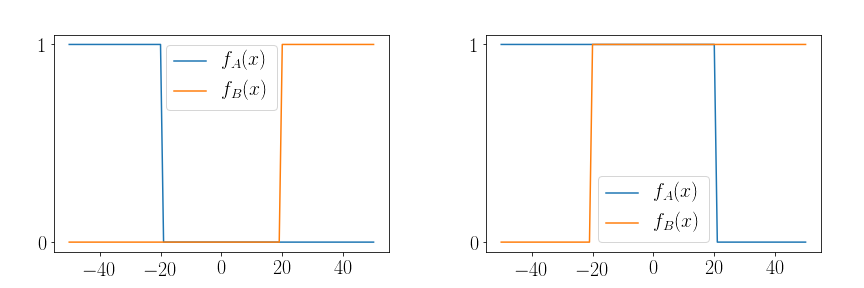}
		
		\caption{Step functions with an interval of a non-biased random walk in the middle, \(d=-20 \) on the left and \(d=20 \) on the right.}
		\label{fig:step-functions}
	\end{figure}

	Fig. \ref{fig:step-functions} shows the plots of these functions for \(d=-20\) and \(d=20\). 
	
Assume now that $d>0$. It is easy to see that it is enough to consider the space state $S' = \{-(d+1),...,d + \tau +1\}$ which is the invariant and attractive set of our dynamics. Stochastic simulations with the above fitness step-functions provide the same qualitative dependence of the expected value of the walker's position on $\tau$ and $d$ as in the hyperbolic-tangent case. Now we calculate the expected value of the walker's position for $d=2$ and $\tau=1$.
	
We are looking for the stationary probability distribution $\pi$ on $S'$. We denote by $\pi^{h}(x,-)$ and $\pi^{h}(x,+)$ stationary probabilities of visiting the state $x$ at time $t$ and respectively the state $x-1$ and $x+1$ at time $t-1$.
	
	Probabilities of states and probabilities of histories of states are related by the following equations:
	\begin{subequations}
		\begin{align}
			\pi(x) &= \pi^{h}(x,-) + \pi^{h}(x,+), -2 \leq x \leq 3, \\
			\pi(-3) &= \pi^{h}(-3,+), \\
			\pi(4) &= \pi^{h}(4,-).
		\end{align}
	\end{subequations}
	
	We solve the above system of equations and get the stationary probability $\pi$ on $S'$ and its expected value $\mu=-0.07$.
	
	In the same way, we derive an analytical formula for the dependence of the expected value of the walker's position on $d$ for both $d>0$ and $d<0$ for $\tau=1$,
	
	\begin{equation}\label{eq:mu-of-d-tau1}
		\mu(d) = \left\lbrace
		\begin{aligned}
			\frac{2d+3}{16d+2} & & \text{ for } & d < -1 \\
			-\frac{1}{16} & & \text{ for } & d = -1 \\
			\frac{1}{13} & & \text{ for } & d = 0 \\
			\frac{1-2d}{16 d+14} & & \text{ for } & d > 0 
		\end{aligned}\right.
	\end{equation}
	In Fig. \ref{fig:tau1} we graph the above function and compare it to simulation results.
	
	\definecolor{darkgreen}{rgb}{0.0, 0.3, 0}
	\begin{figure}
		\centering
		\includegraphics[width=\linewidth]{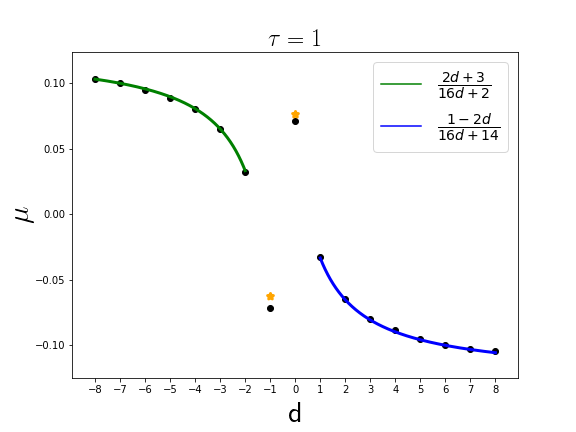}
		
		\caption{Analytical dependence $\mu(d)$  and results of simulations for $\tau=1$. Black points $\bullet$ represent data points obtained in simulation. {\color{darkgreen} Green line} represents the formula in eq. \ref{eq:mu-of-d-tau1} for $d < -1$. {\color{orange} Orange stars  $\star$} represent values in the formula in eq. \ref{eq:mu-of-d-tau1} for  $d = -1$ and $d = 0$. {\color{blue} Blue  line} represents the formula in eq. \ref{eq:mu-of-d-tau1} for $d > 0$. }
		\label{fig:tau1}
	\end{figure}
	
	{\bf Discussion.} We studied random walks with asymmetric time delays. We showed a novel behavior --- a shift in the expected value of the position of the walker as a joint effect of both time delays and stochasticity. We showed that a mean position of the walker is a monotonic function of one delay (we set the other one to 0).
	
	If both fitness functions are almost (or exactly in the step-function case) equal to $1$ on the interval of the length $2d$ around the origin, then the expected value of the walker is negative and a decreasing function of the time delay, a linear one for small time delays, and a decreasing function of $d$. However, if both fitness functions are almost equal to $0$ on this interval, then the effect of time delays is reversed --- the expected value of the walker is positive, an increasing function of both the time delay and $|d|.$
	
	Such an effect cannot be achieved in our models by time delays or stochasticity alone. In the deterministic version of our model, there appear symmetric cycles around the stationary point so the mean position of the walker stays the same. In the stochastic version without time delays, the expected value of the position of the walker is given by the stationary point.
	
	One can interpret our models as discrete replicator dynamics with two strategies $A$ and $B$, fitness functions given by $f_{A}$ and $f_{B}$, and with strategy-dependent time delays. The correspondence is exact in models with a finite state space as in our models with fitness step functions. Our results show that the effects of strategy-dependent time delays on the equilibrium structure of populations may depend on details of fitness functions. 

	\begin{acknowledgments} 
		We would like to thank the National Science Centre (Poland) for financial support under the grant 2015/17/B/ST1/00693.
	\end{acknowledgments}

\end{document}